# Modelling MEMS structures using Cosserat theory

M. Calis[1,3], O. Lagrouche[2], M.P.Y Desmulliez[3]

[1]Corac Group plc
Brunel Science Park
Uxbridge, UB8 3PQ

School of the Built Environment
Heriot-Watt University
Edinburgh, EH14 4AS, UK

[3] School of Engineering & Physical Sciences
MIcroSystems Engineering Center (MISEC)
Heriot-Watt University, Eindburgh, EH14 4AS

**Modelling MEMS involves a variety of software tools that deal with the analysis of complex geometrical structures and the assessment of various interactions among different energy domains and components. Moreover, the MEMS market is growing very fast, but surprisingly, there is a paucity of modelling and simulation methodology for precise performance verification of MEMS products in the nonlinear regime. For that reason, an efficient and rapid modelling approach is proposed that meets the linear and nonlinear dynamic behaviour of MEMS systems.**

## I. Introduction

At the present time, modelling MEMS is widely made using either Finite Element Analysis (FEA) [1], component level modelling [2][3], Boundary-Element Method (BEM) [4] or lumped level modelling [5]. The main disadvantage of the FEA is the dependency of stress on the number of elements used to represent a component such as a microbeam cantilever. For accurate representation, the number of equations increases significantly and the model tends to be cumbersome and complicated, preventing thereby designers from performing real-time simulation. Even with the complexity involved by the FEA approach, this technique is unsuitable for taking into account large scale deformation/motion. Component level (also call nodal) methods have been introduced in the form of MEMS design tools containing a library parameterised behavioural models. This tool describes each component as a single element enabling a considerable reduction in simulation time compared to finite element models where the component is often subdivided into many elements. A major disadvantage of this component level method is the limitation of its library containing only frequently used MEMS components. In lumped level modelling the simulation time is also low since only the behaviour of the transducer is modelled. BEM requires only surface discretization and the treatment of boundary conditions at infinity, however discretizing the boundary integral equations leads to cumbersome systems whose memory costs scale as $O(n^2)$ and solution costs scale with $O(n^3)$ with $n$ being the number of discretization unknowns [6].

A new approach is therefore unavoidable to reduce the design process and to enable simulation of complex MEMS structures. In that respect, this paper presents a new approach for modelling linear and especially nonlinear MEMS structures based on Cosserat theory for a better representation of stress in miniaturized systems, especially in the nonlinear regime. The use of Cosserat theory leads to a reduction of the complexity of the modelling and thus increases its capability to simulate microstructures in real-time, indispensable for haptic technology. To demonstrate the feasibility of the proposed model, a cantilever microbeam undergoing loads modelled with ANSYS, SABER and Cosserat theory are compared.

## II. Methods for Modelling Mems Components

Methods for modelling MEMS components can usually be classified into two categories as shown in Figure 1. The exact methods include the Euler-Bernoulli [7] and the Timoshenko techniques [8] which are solved using a power series expansion. The FEA, BEM and lumped mass methods are classified under the approximate methods and are solved using superposition techniques. The FEM is also known as the matrix displacement method. The word approximate is used since it assumes that displacements can be represented by simple polynomial expressions. Our approach uses a semi-analytical method based on both power series expansions and a multimodal approximate method.

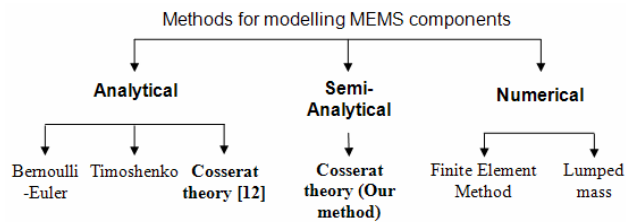

Fig. 1. Taxonomy of MEMS modelling Methods

MEMS devices consist of a system of inter-connected slender structures and masses which is also known as a network of slenders. The Cosserat theory can implement such modelling systems as a network of Cosserat rods where each Cosserat rod will be further discretised in order to yield a dynamic description using ordinary differential equations (ODEs) instead of partial differential equations (PDEs). This considerably facilitates the use of numerical and analytical techniques; therefore it is possible to implement it using VHDL-AMS.

## III. Cosserat Theory

A Cosserat rod can be described by defining a set of cross-sections the centroids of which are connected by a curve which is referred to as the line of centroids.

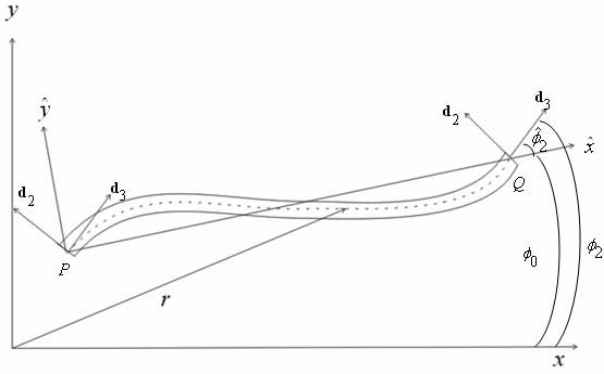

Fig. 2. A Cosserat Rod

Figure 2 is a schematic construction of a 2D Cosserat beam element in the $x$ - and $y$ - plane. The motion in space of a nonlinear Cosserat rod segment can be represented as a vector $r(s,t)$, called a Cosserat curve, which describes the position of the line of centroids of the cross-sections (Figure 2, dotted line). The deformation of the slender MEMS structure represented by the deformation of the centroid line depends upon three vectors $\mathbf{r}(s)$, $\mathbf{d}_1(s)$, and $\mathbf{d}_2(s)$. The modelling of the microstructures is based on the centroid line and a director replacing a detailed 3D meshing used in FEM. Figure 2 represents a schematic 2D Cosserat beam element in a moving frame where $P$ and $Q$ are two time-dependent end points having inertial Cartesian coordinates $(x_1(t), y_1(t))$ and $(x_2(t), y_2(t))$. $\phi_1(t)$ and $\phi_2(t)$ denote the angles between $d_3$ and $e_1$ at node $P$ and $Q$, respectively. The shape function of the rod can be constructed by approximating PDEs with ODEs and interpolating $r(s,t)$ for given end point values of the beams, namely $x_1(t)$, $y_1(t)$, $\phi_1(t)$ at $P$ and $x_2(t)$, $y_2(t)$, $\phi_2(t)$ at $Q$.

In the Cosserat theory, the accuracy will depend on the method used to model the motion/deformation of the centroid line. Unlike in [9] where the Newton's dynamical law and analytical method are used, our approach is based on a semi-analytical method and on the Euler-Bernoulli equation of motion.

$$c^4 \frac{\partial^4 u_y}{\partial x^4} + \ddot{u}_y = 0 \qquad (1)$$

where $c^4 = \dfrac{EI}{\rho A}$, $I$ is the moment of inertia of the beam of cross-section $A$ and Young modulus $E$. To solve the Euler-Bernoulli equation of motion, the displacements $u_x$ and $u_y$ in the transverse and axial directions are expanded in ascending powers of $w$ [10].

$$\mathbf{u} = \begin{bmatrix} u_x \\ u_y \end{bmatrix} = \left( \begin{bmatrix} \mathbf{a}_{0x} \\ \mathbf{a}_{0y} \end{bmatrix} + w \begin{bmatrix} \mathbf{a}_{1x} \\ \mathbf{a}_{1y} \end{bmatrix} + w^2 \begin{bmatrix} \mathbf{a}_{2x} \\ \mathbf{a}_{2y} \end{bmatrix} + w^3 \begin{bmatrix} \mathbf{a}_{3x} \\ \mathbf{a}_{3y} \end{bmatrix} + .... \right)$$

$$\qquad (2)$$

$$= \left( \mathbf{a}_0 + w\mathbf{a}_1 + w^2\mathbf{a}_2 + w^3\mathbf{a}_3 + ... \right) \mathbf{U} = \mathbf{a}\mathbf{q}e^{iwt} \qquad (3)$$

where

$$\mathbf{U} = \{U_1 \quad U_2 \quad U_3 \quad U_4\} = \{q_1 \quad q_2 \quad q_3 \quad q_4\} e^{iwt} \quad (4)$$

The matrices $\mathbf{a}_{ix}$ and $\mathbf{a}_{iy}$ represent axial and transverse displacements, respectively. $w$ is the circular frequency. In equation (2) $u_x$ and $u_y$ can be rewritten respectively as

$$u_x = \sum_{r=0}^{\infty} w^r \mathbf{a}_{rx}\mathbf{q}e^{iwt} = \mathbf{a}_x\mathbf{q}e^{iwt} \qquad (5)$$

$$u_y = \sum_{r=0}^{\infty} w^r \mathbf{a}_{ry}\mathbf{q}e^{iwt} = \mathbf{a}_y\mathbf{q}e^{iwt} \qquad (6)$$

Substituting equations (5) and (6) into equation (1) we get the following equation

$$c^4 \sum_{r=0}^{\infty} w^r a_{ry}^{iv} q e^{iwt} - w^2 \sum_{r=0}^{\infty} w^r a_{ry} q e^{iwt} = 0 \qquad (7)$$

As we are only considering geometric nonlinearity, the matrix $\mathbf{a}_{0y}$ is determined from the boundary conditions of the cantilever microbeam.

## IV. RESULTS

To validate our design approach, our model is tested on a 2D microbeam cantilever undergoing transversal loads applied at the free end. Figure 3 shows the deflections for a given load obtained from the analytical solution, ANSYS and Cosserat theory using linear stiffness matrix.

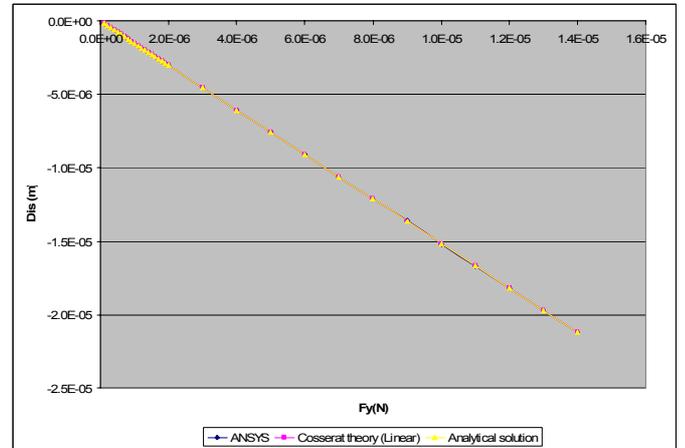

Fig. 3. Deflection of 2d cantilever microbeam

From the figure above, the results obtained for the different model are identical. Secondly, our model is tested in the same conditions as in [11]. The test consists of a static behaviour of a cantilever microbeam where a force of $7.3*10^{-4} N$ was applied at the free end in $x$ and $y$ directions. Table I compares the results of the analytical calculations, static analysis of the ANSYS and SABER models [11], with our model. Afterwards, to validate the nonlinear model, we compare the analytical buckling load [12] against the buckling load obtained using our model.

TABLE I
STATIC ANALYSIS OF THE LINEAR CANTILEVER MICROBEAM

| Static Load | Analytical solution | ANSYS (100 elements) [11] | SABER (2 elements) [11] | SABER (8 elements) [11] | Our model (2 elements) | Our model (8 elements) |
|---|---|---|---|---|---|---|
| $F_z = 7.3*10^{-4}$ | 2.296 | 2.298 | 2.296 | 2.296 | 2.296 | 2.296 |
| $F_x = 7.3*10^{-4}$ | $8.971*10^{-3}$ | $8.971*10^{-3}$ | $8.971*10^{-3}$ | $8.971*10^{-3}$ | $8.971*10^{-3}$ | $8.971*10^{-3}$ |

TABLE II
BUCKLING EFFECT

| | Analytical solution [12] | Our model | Error (%) |
|---|---|---|---|
| 1 CRE | $5.42828*10^{-5}$ | $5.46911*10^{-5}$ | 0.746616 |
| 4 CREs | $5.42828*10^{-5}$ | $5.42846*10^{-5}$ | 0.003276 |
| 10 CREs | $5.42828*10^{-5}$ | $5.42825*10^{-5}$ | 0.000055 |

The results show clearly that model is in good agreement with the SABER, ANSYS and analytical solutions (Table I). Then the nonlinear model is used to compute the first buckling load of the microbeam. Table II shows that the results obtained for $1$, $4$ and $10$ elements for the discretization of the microbeam are very close to the analytical solution. Currently the nonlinear model is being used to compute the deflection of a cantilever microbeam when it is subjected to a vertical load at the free-end. Such a problem is not straightforward and extra physical considerations must be introduced in order to get the simulation close to the reality.

## IV. CONCLUSION

In this paper, it is demonstrated that the proposed method for modelling linear effects in MEMS is valid. The Cosserat theory has been successfully used for modelling and testing simple structures such as a cantilever and a microbridge. Nonlinear problems such as the buckling of beams have also been tackled successfully. We believe that Cosserat theory will lead to a reduction of the complexity of the modelling and thus will increase its capability for real time simulation, indispensable for haptic technologies.